\documentclass{aastex63}

\usepackage{times}
\usepackage{amsmath,amssymb,amstext}
\hypersetup{linkcolor=red,citecolor=blue,filecolor=cyan,urlcolor=blue}
\usepackage{tabularx}
\usepackage{longtable}
\usepackage{float}
\usepackage{natbib}
\usepackage{graphicx,color}
\usepackage{txfonts}

\usepackage{multirow}
\usepackage{array}
\usepackage{rotating}
\usepackage{sidecap}
\usepackage{hyperref}
\usepackage{epstopdf}
\usepackage{footnote}
\usepackage{tabularx}
\usepackage{booktabs}
\usepackage{longtable}
\usepackage{tabu}
\usepackage{longtable}

\newcommand{\kms}{km\,s$^{-1}$}
\def\eg{{\it e.g.,\ }}

\begin{document}

\title{{\large{\bf Plasmoids, Flows, and Jets During Magnetic Reconnection in a Failed Solar Eruption}}}
\author{Pankaj Kumar\altaffiliation{1,2}}

\affiliation{Department of Physics, American University, Washington, DC 20016, USA}
\affiliation{Heliophysics Science Division, NASA Goddard Space Flight Center, Greenbelt, MD, 20771, USA}

\author{Judith T.\ Karpen}
\affiliation{Heliophysics Science Division, NASA Goddard Space Flight Center, Greenbelt, MD, 20771, USA}

\author{Spiro K. Antiochos}
\affiliation{CLaSP, University of Michigan, Ann Arbor, MI, 48109, USA }

\author{C.\ Richard DeVore}
\affiliation{Heliophysics Science Division, NASA Goddard Space Flight Center, Greenbelt, MD, 20771, USA}

\author{Peter F.\ Wyper}
\affiliation{Department of Mathematical Sciences, Durham University, Durham DH1 3LE, UK}

\author{Kyung-Suk Cho}
\affiliation{Korea Astronomy and Space Science Institute, Daejeon, 305-348, Korea}
\affiliation{University of Science and Technology, Daejeon 305-348, Korea}

\email{pankaj.kumar@nasa.gov}

\begin{abstract}

 {We report a detailed analysis of a failed eruption and flare in active region 12018 on 2014 April 3 using multiwavelength observations from SDO/AIA, IRIS, STEREO, and Hinode/SOT. At least four jets were observed to emanate from the cusp of this small active region (large bright point) with a null-point topology during the two hours prior to the slow rise of a filament. During the filament slow rise multiple plasma blobs were seen, most likely formed in a null-point current sheet near the cusp. The subsequent filament eruption, which was outside the IRIS field of view, was accompanied by a flare but remained confined. During the explosive flare reconnection phase, additional blobs appeared repetitively and moved bidirectionally 
 within the flaring region below the erupting filament. The filament kinked, rotated, and underwent leg-leg reconnection as it rose, yet it failed to produce a coronal mass ejection. Tiny jet-like features in the fan loops were detected during the filament slow-rise/pre-flare phase. We interpret them as signatures of reconnection between the ambient magnetic field and the plasmoids leaving the null-point sheet and streaming along the fan loops. We contrast our interpretation of these tiny jets, which occur within the large-scale context of a failed filament eruption, with the local nanoflare-heating scenario proposed by \citet{antolin2021}.}

\end{abstract}
\keywords{Sun: jets---Sun: corona---Sun: UV radiation---Sun: magnetic fields}

\begin{figure*}
\centering{
\includegraphics[width=18cm]{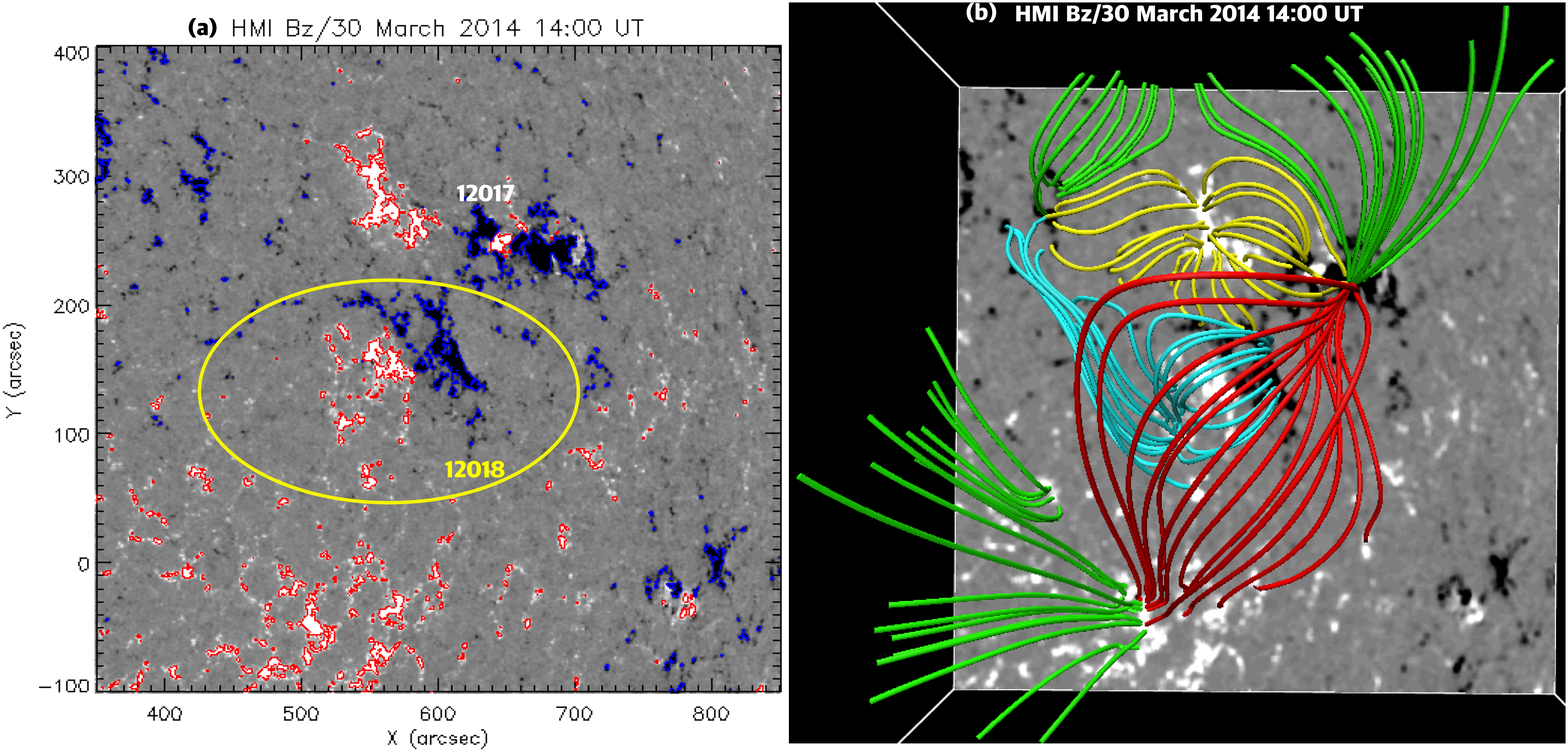}
\includegraphics[width=18cm]{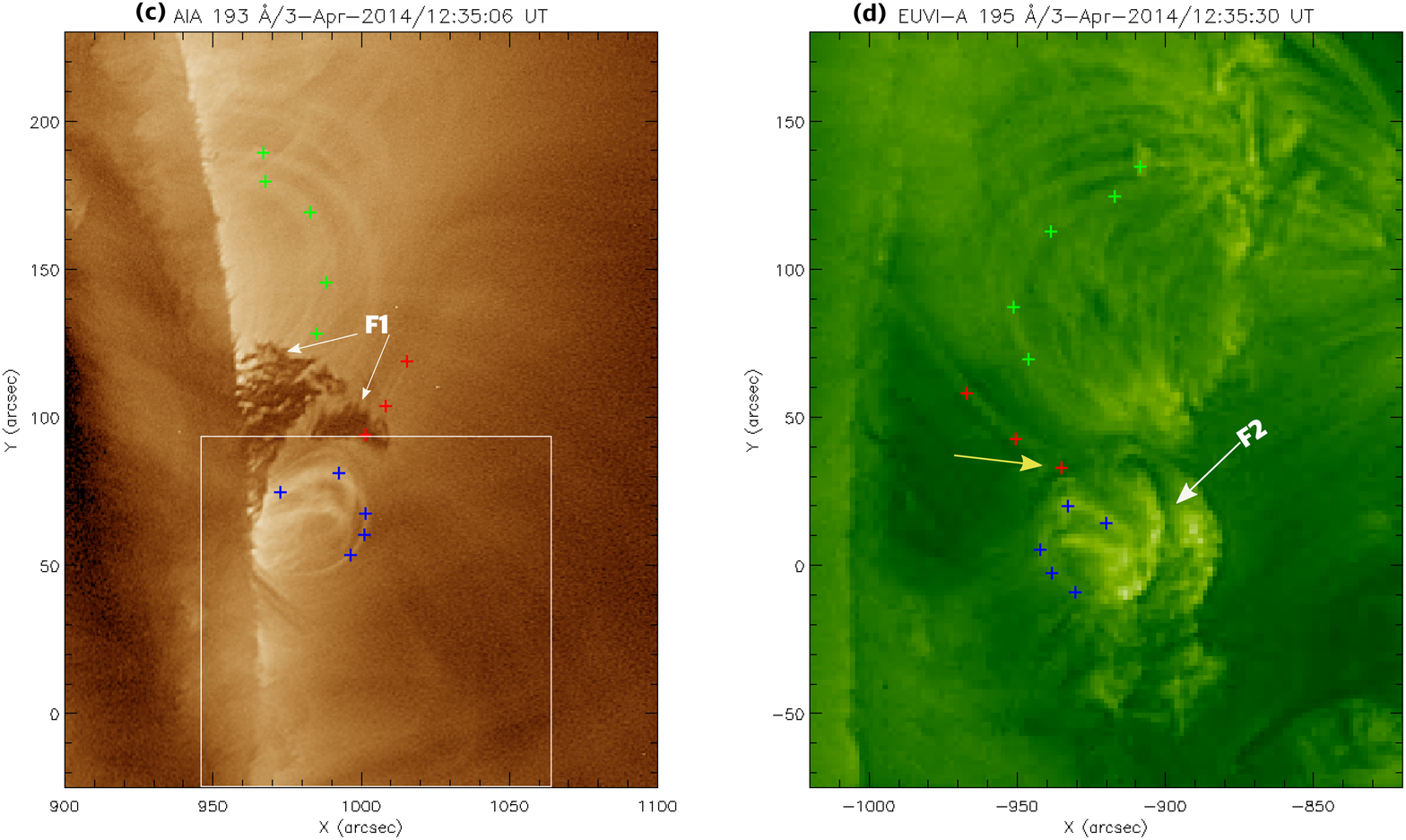}
}
\caption{ (a) SDO/HMI magnetogram showing ARs 12017 and 12018 on 2014 March 30. Red/blue contours indicate positive/negative polarities (levels = $\pm$100 G). The filament eruption took place in AR 12018 (outlined by the yellow ellipse). (b) Potential-field extrapolation of the ARs (same field of view as shown in panel (a)). Red field lines are arcades overlying AR 12018 (cyan). Open field lines are shown in green. Yellow field lines represent connecting loops in the neighbouring AR 12017 (from \citealt{kumar2022}). (c, d) Cotemporal SDO/AIA 193 {\AA} and STEREO/EUVI-A 195 {\AA} images of AR 12018.  The white box outlines the IRIS field of view. F1 and F2 are two filaments located along the PIL within AR 12018. The plus symbols are the data points used to identify the same structures in SDO and STEREO images. The yellow arrow points to the apparent location of a cusp, which we interpret as the site of a null point.} 
\label{fig1}
\end{figure*}

\section{INTRODUCTION}\label{intro}
Coronal bright points are the main sources of coronal jets on a wide range of scales, from barely resolved features in coronal holes and quiet-Sun to small active regions \citep{raouafi2016}. These jets are often associated with eruptions from a filament channel, with or without a filament \citep{sterling2015, kumar2019a}. The magnetic structure of most bright points is that of an embedded bipole, consisting of a fan of loops bounded by a dome-shaped separatrix surmounted by a null point. A spine line passes through the null, leading to the term ``fan-spine'' topology for this configuration \citep{shibata1994,priest1996}. Slow reconnection between the closed field above the filament channel and the ambient open field, at the current sheet created by deformation of the null, generates faint jets along with opening of the field lines. This slow reconnection associated with the null is a key feature of the breakout model for coronal eruptions on all scales \citep{antiochos1999,wyper2017}, and is hereafter denoted breakout reconnection at the breakout current sheet (BCS).  Reconnection at the BCS has been inferred from EUV observations \citep{kumar2018,kumar2019b,kumar2021} and directly observed in numerical magnetohydrodynamics (MHD) simulations \citep{wyper2017,wyper2018,wyper2021} of coronal jets. 
Multiple blobs and bidirectional jets in the BCS have been observed in {\it Solar Dynamics Observatory}/Atmospheric Imaging Assembly (SDO/AIA; \citealt{lemen2012}) and Interface Region Imaging Spectrograph (IRIS; \citealt{bart2014}) images before the onset of explosive flare reconnection \citep{kumar2019b}. These blobs, which are commonly interpreted as plasmoids or magnetic islands, initially appear near the main BCS null, propagate along the BCS, and continue onto the separatrix dome. Magnetic reconnection at a flare current sheet (FCS) that forms underneath the rising filament builds the encompassing flux rope. The onset of fast flare reconnection is often associated with bidirectionally traveling blobs in the FCS, interpreted as plasmoids. Explosive breakout reconnection destroys the flux rope when it encounters the BCS and produces untwisting jets \citep{wyper2017,wyper2018,kumar2018}. Before and after eruptive activity, these null-point topologies often produce cool coronal rain emanating from the null point, most likely due to slow interchange reconnection \citep{mason2019, kumar2019a, kumar2021}.
Although a continuum of eruptive events (jets to coronal mass ejections) has been detected from bipolar coronal bright points to complex, multipolar active regions (ARs), it is still unclear why some eruptions are successful while others remain confined. Therefore, understanding  failed eruptions is essential to understand the overall eruption process. 

We analyzed in depth multi-spacecraft observations of a jet followed by a failed eruption from a small active region, NOAA 12018, on 2014 April 3. 
Our comprehensive analysis of this event utilized data from the SDO/AIA, IRIS, and {\it Solar TErrestrial RElations Observatory} (STEREO) SECCHI-A (Ahead; \citealt{wuelser2004,howard2008}) missions, providing both high-resolution multithermal coronal imaging over large fields of view (FOVs) and higher-resolution imaging of cooler plasma within a limited FOV. \citet{antolin2021} reported tiny jet-like features along and transverse to coronal loops during the early phase of the filament eruption, and interpreted them as nanojets associated with nanoflares \citep{parker1988,klimchuk2015}.  
Their study utilized only the IRIS and reduced AIA fields of view encompassing the southern portion of the configuration. In contrast, we conclude that the flows, heating, and plasmoids detected in the EUV were generated by the failed filament eruption that produced a flare without a coronal mass ejection (CME). 


\begin{figure*}
\centering{
\includegraphics[width=16cm]{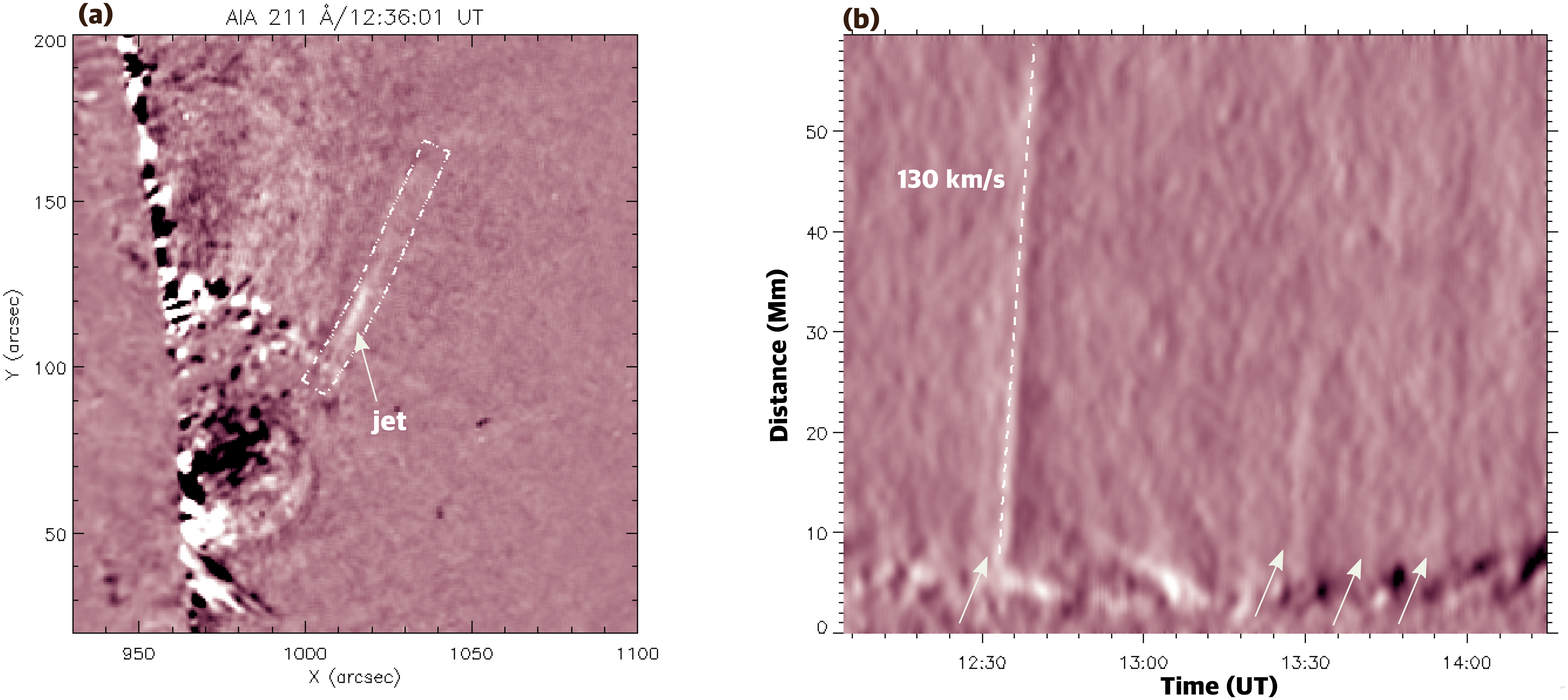}
\includegraphics[width=16cm]{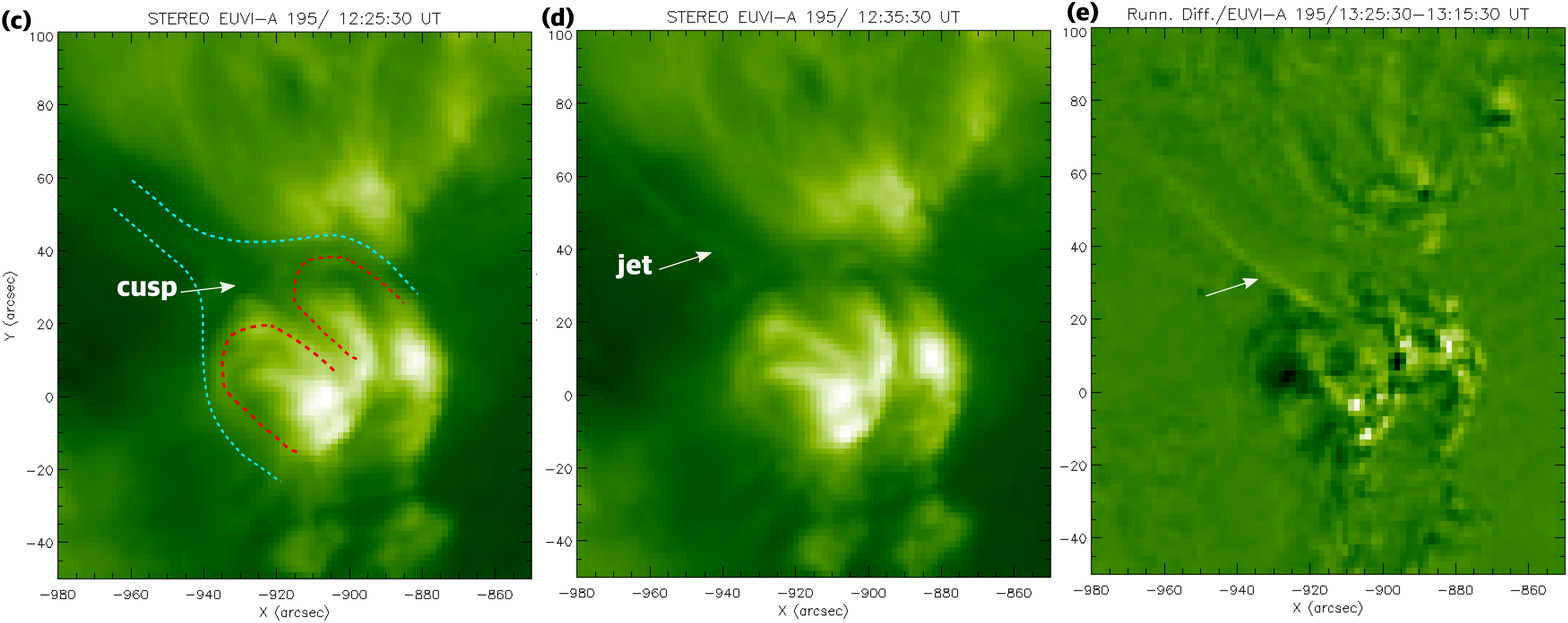}

}
\caption{(a) AIA 211 {\AA} running-difference image ($\Delta t$ = 2 min) with a pre-eruption jet marked by an arrow. The rectangular box outlines the slit used to produce panel (b). (b) Time-distance running-difference intensity plot in the AIA 211 {\AA} channel along the direction of jet propagation. Arrows indicate multiple pre-eruption jets. (c,d) EUVI-A 195 {\AA} intensity and (e) running-difference images showing the cusp and two pre-eruption jets. An animation of the (a) and (b) panels is available. The animation runs from 12:05:13-14:15:37 UT. (An animation of this Figure is available online.)} 
\label{fig2}
\end{figure*}

\begin{figure*}
\centering{
\includegraphics[width=18cm]{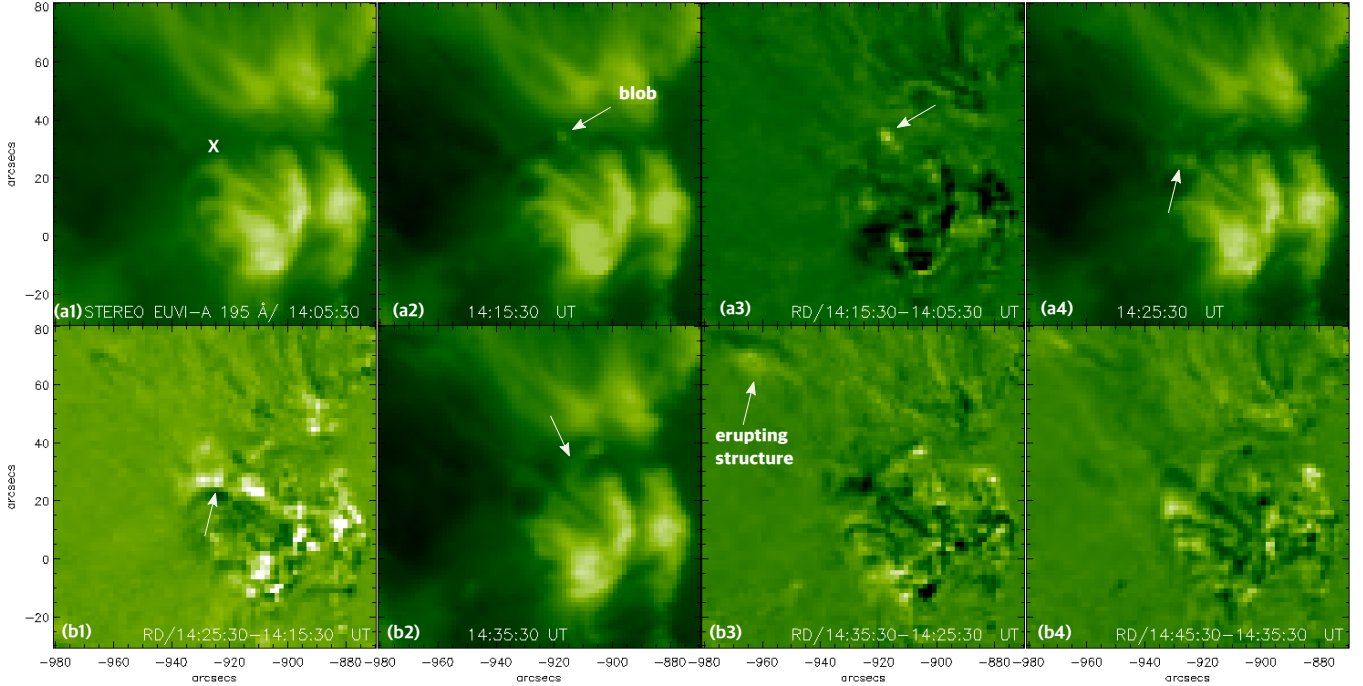}
}
\caption{ STEREO/EUVI-A 195 {\AA} intensity images (a1, a2, a4, and b2) and running-difference images (a3, b1, b3, and b4) during the filament slow-rise phase. Arrows point to  outflows near the null at the position marked by an X in panel (a1). An animation of this Figure is available. The animation runs from 12:15:30-16:25:30 UT. (An animation of this Figure is available online.)
} 
\label{fig3}
\end{figure*}

\begin{figure*}
\centering{
\includegraphics[width=18cm]{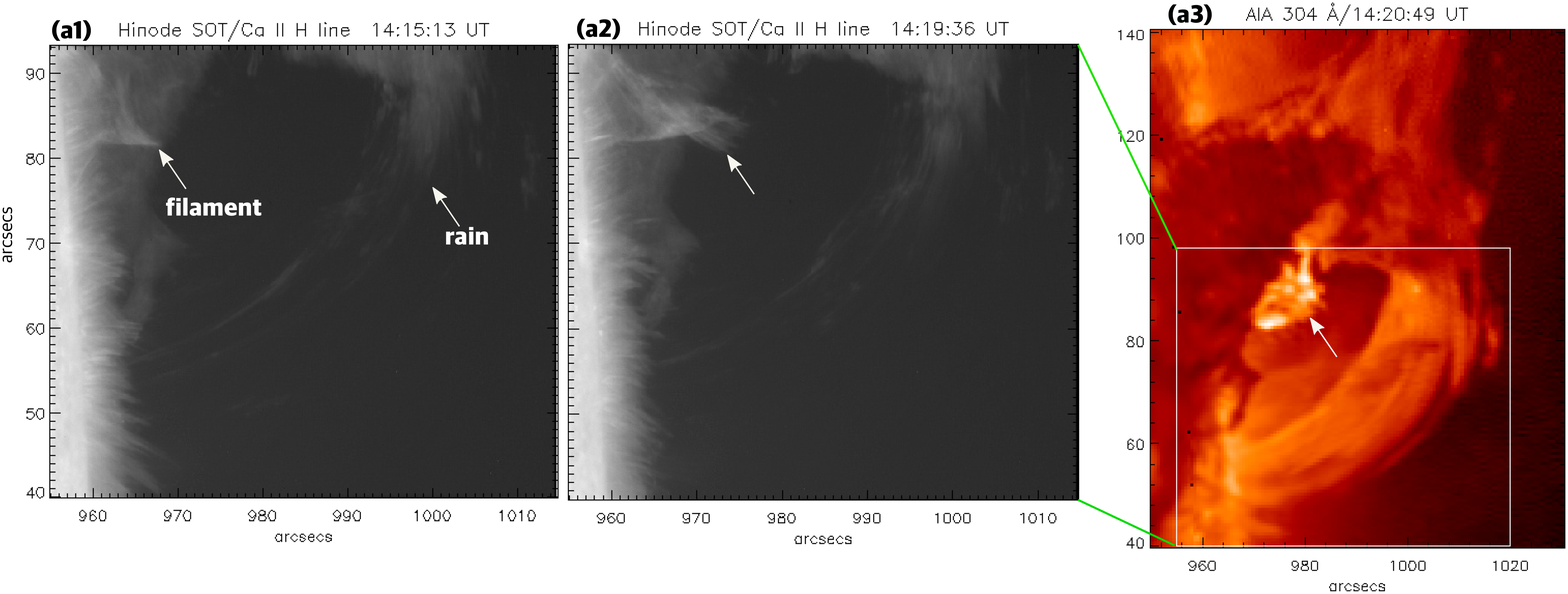}
\includegraphics[width=18cm]{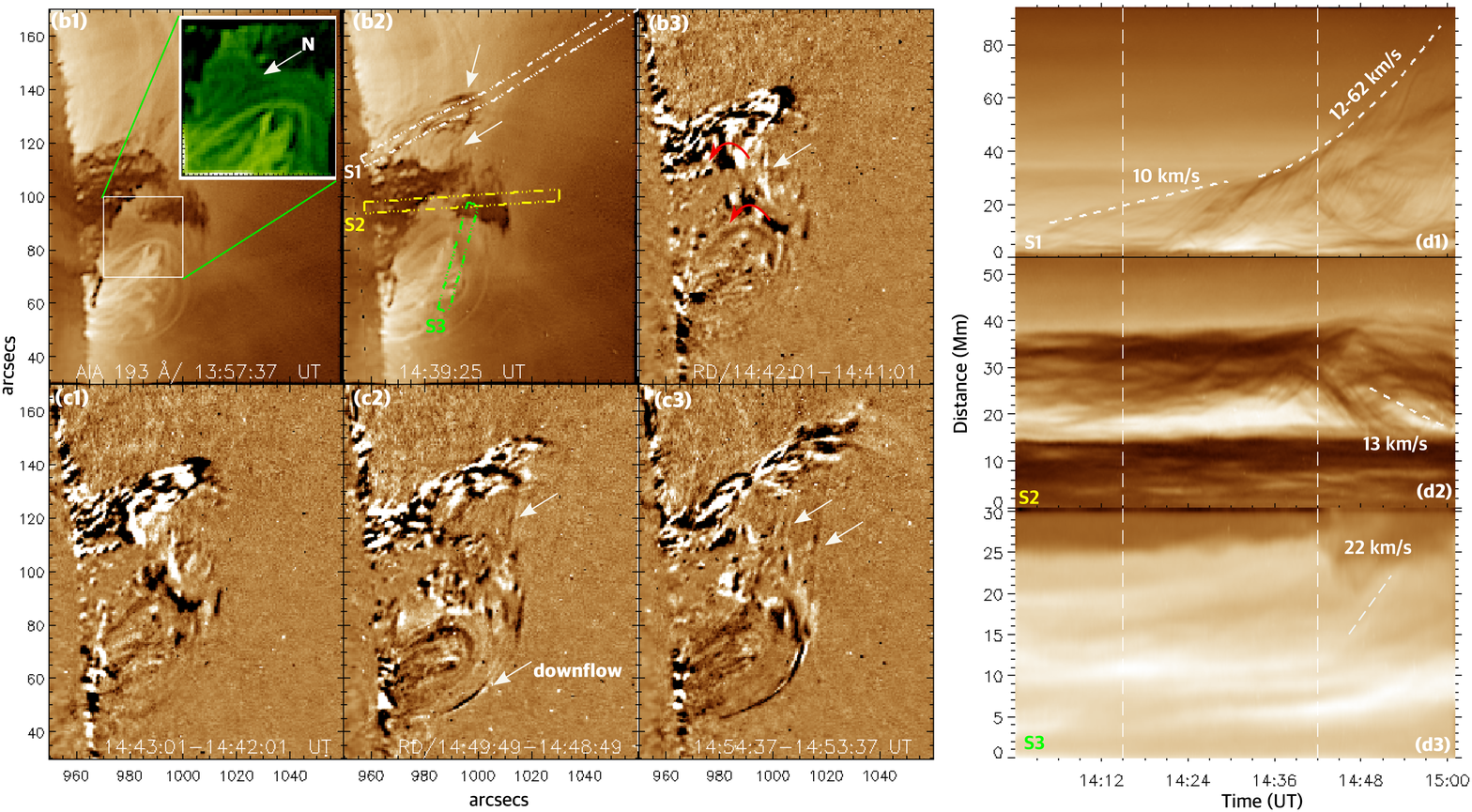}
}
\caption{(a1,a2) Selected Hinode/SOT Ca II H-line images showing the filament activation and coronal rain. (a3) AIA 304 {\AA} image with larger FOV. The white box indicates the SOT field of view. (b1,b2) AIA 193 {\AA} images showing the cusp (N) and rising filament F1 (marked by arrows). (b3, c1,c2,c3) AIA 193 {\AA} running-difference images of the erupting filament.  Two red arrows indicate the rotating threads of the filament. White arrows show the threads connecting F1 to the southern structure near the cusp. (d1-d3) Time-distance intensity maps along slits S1, S2, and S3 marked in panel b2. The first vertical dashed line at 14:15 UT indicates the filament activation onset time derived from the SOT images. The second vertical dashed line at 14:42 UT shows the rotation onset time of the filament’s southern segment. Distance scale starts at the leftmost ends of the slits shown in (b2). An animation of the panels (b1)-(d3) is available. The first part of the animation runs from 13:02:49-15:57:13 UT and second part runs from 14:00:01-15:00:01 UT.  (An animation of this Figure is available online.)
} 
\label{fig4}
\end{figure*}
\begin{figure*}
\centering{
\includegraphics[width=18cm]{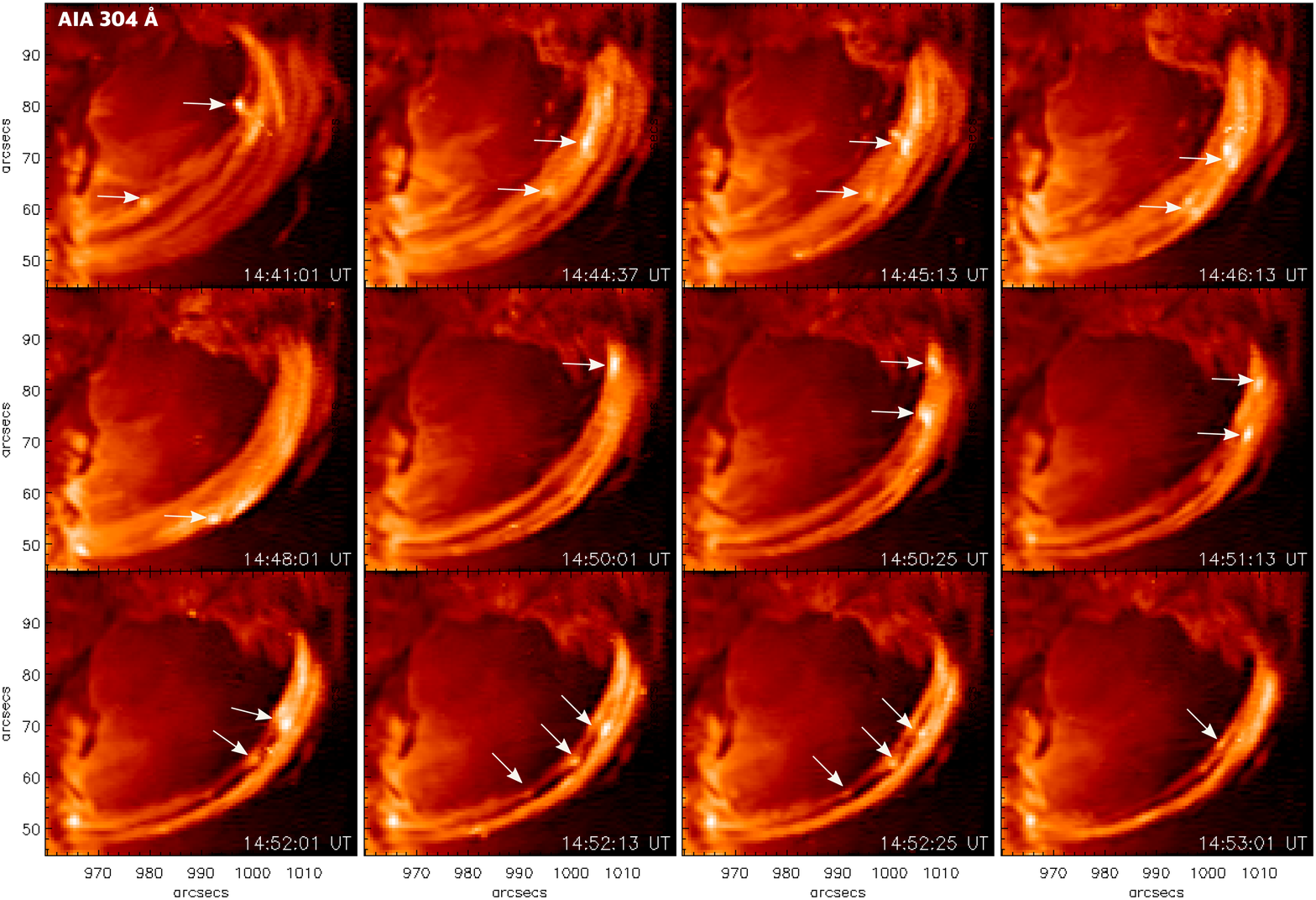}
\includegraphics[width=18cm]{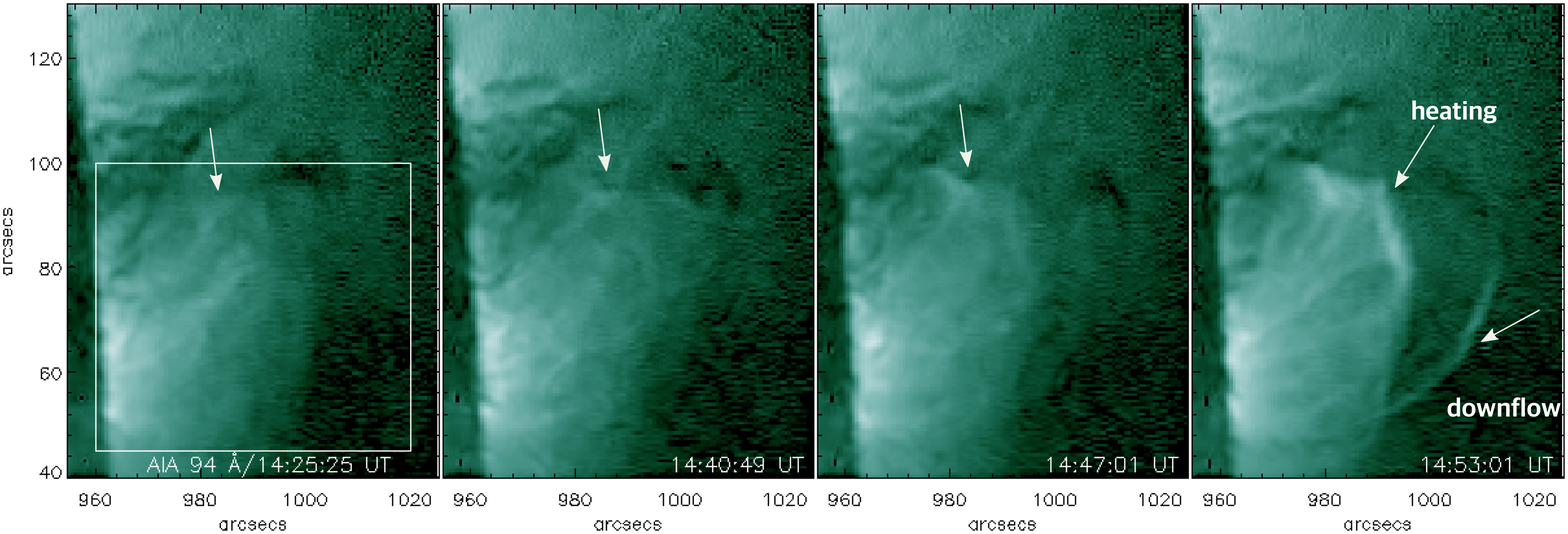}
}
\caption{ Top: Selected AIA 304 {\AA} images showing multiple plasma blobs (marked by arrows) and coronal rain during 14:41-14:53 UT. Bottom: AIA 94 {\AA} images showing flows and heating at and below the BCS, marked by arrows, during 14:25:25-14:53:01 UT. The white box indicates the AIA 304 {\AA} FOV shown in the top panels. An animation of this Figure is available. The first part (AIA 304 \AA) of the animation runs from 13:01:37-15:57:25 UT and second part (AIA 94 \AA) runs from 13:01:49-15:57:25 UT. (An animation of this Figure is available online.)} 
\label{fig5}
\end{figure*}
\begin{figure*}
\centering{
\includegraphics[width=16cm]{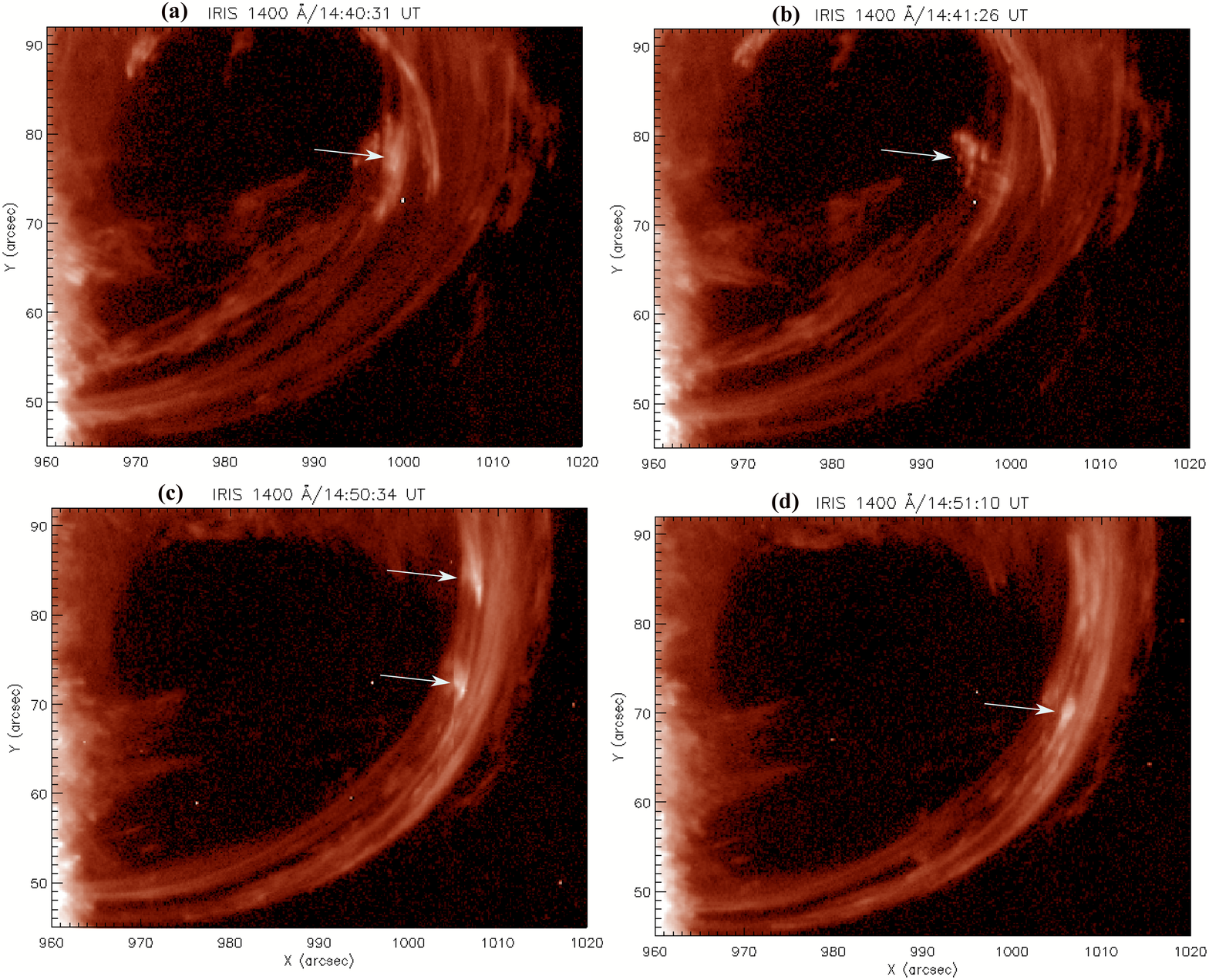}
\includegraphics[width=16cm]{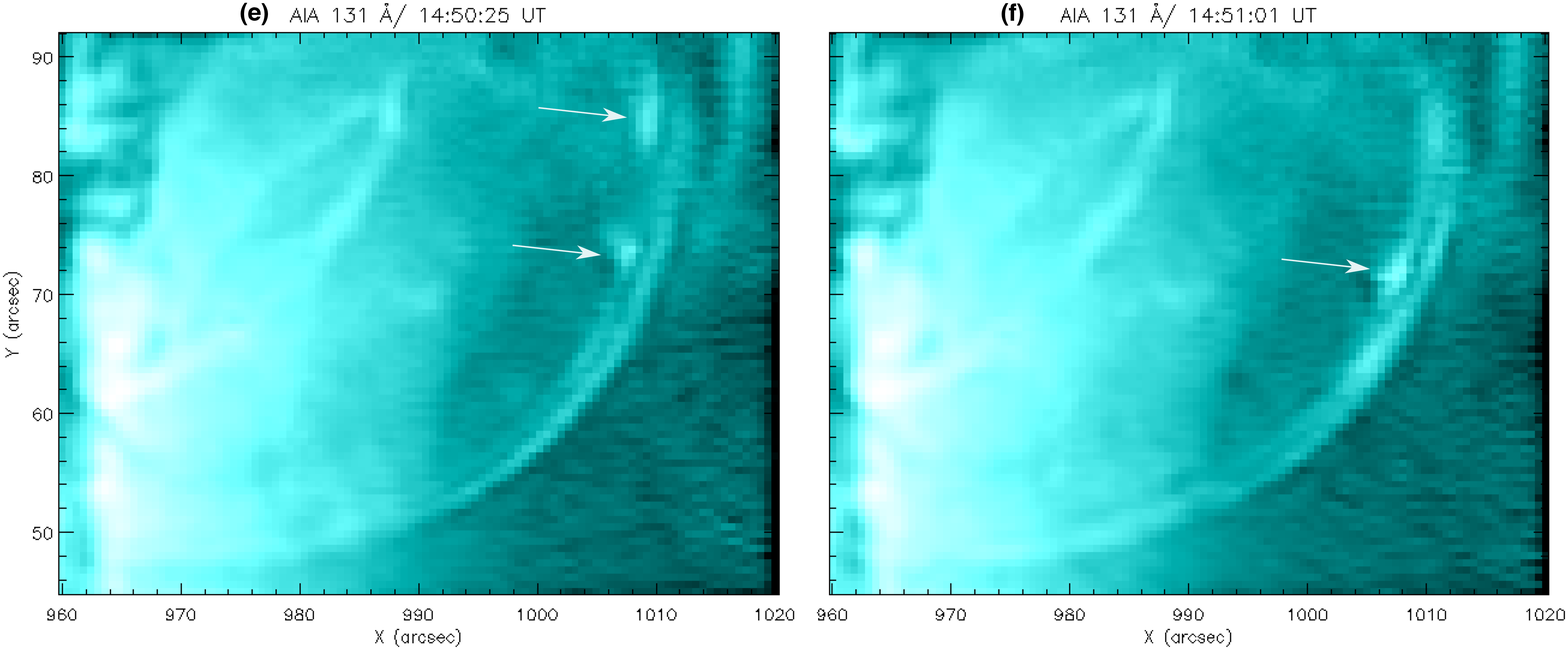}
}
\caption{ (a-d) Selected IRIS 1400 {\AA} images showing multiple plasma blobs (marked by arrows) and coronal rain. (e-f) AIA 131 {\AA} images showing plasma blobs (marked by arrows) and brightenings near the footpoint.} 
\label{fig6}
\end{figure*}
\begin{figure*}
\centering{
\includegraphics[width=16.5cm]{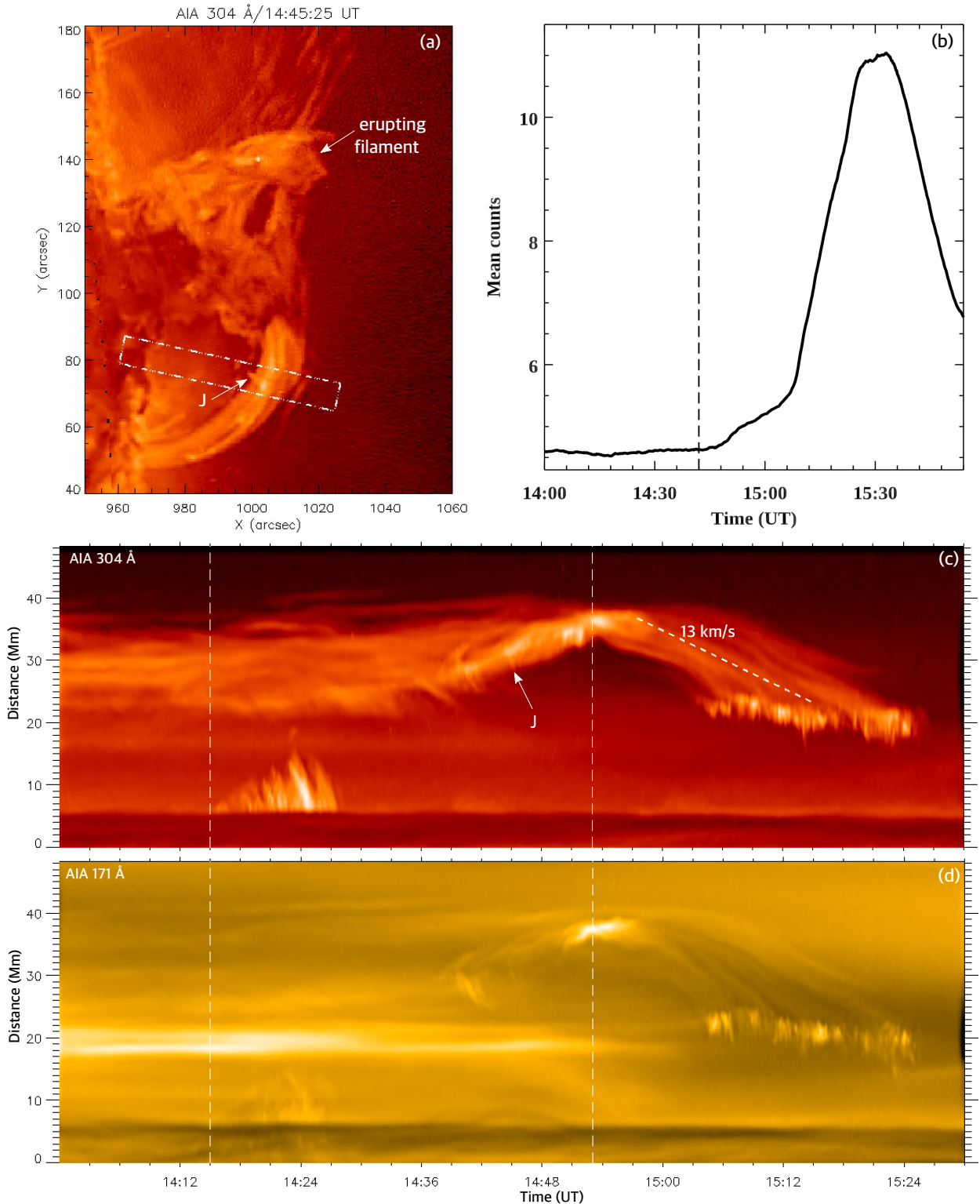}
}
\caption{(a) AIA 304 {\AA} image during the failed filament eruption. J is the jet-like feature associated with a large blob. The dashed-dotted white rectangle outlines the slice used to produce the TD plots in (c) and (d). (b) AIA 94 {\AA} mean counts (arbitrary unit) from the flare region (bottom-left panel in Figure \ref{fig5}). A vertical dashed line at 14:42 UT indicates the flare onset. (c) AIA 304 {\AA} and (d) 171 {\AA} TD intensity plots along the slice during 14:00-15:28 UT. Vertical dashed lines mark the onset of filament activation (14:15 UT) and the start of the loop contraction (14:53 UT). Distance scale starts at the left end of the slit shown in (a). 
} 
\label{fig7}
\end{figure*}
\begin{figure*}
\centering{
\includegraphics[width=18cm]{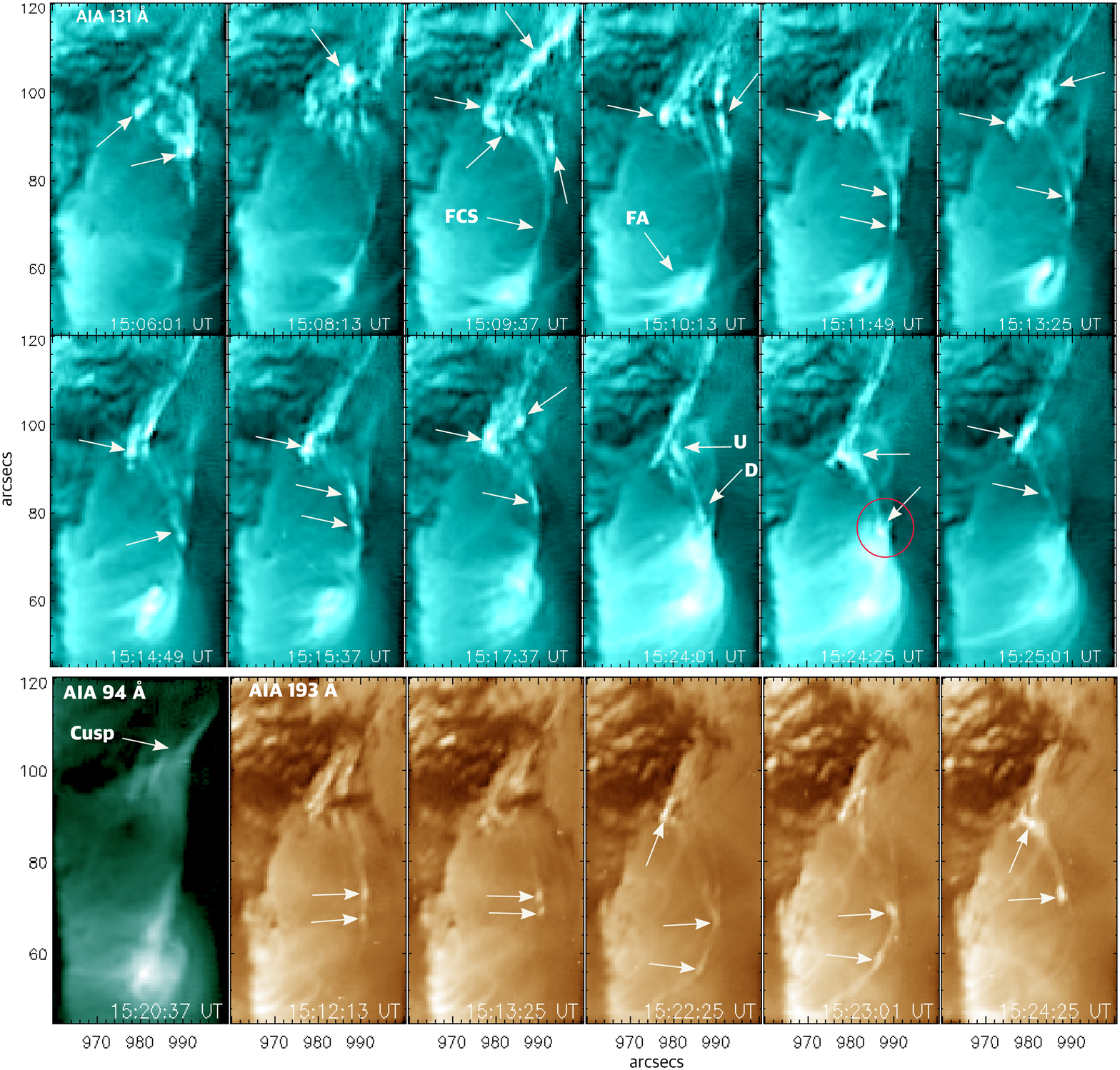}
}
\caption{AIA 131 and 193 {\AA} images during the flare (15:06-15:25 UT). Arrows indicate multiple bright blobs in the flare current sheet (FCS). AIA 94 {\AA} (bottom left) shows a cusp. FA=flare arcade, U=upward moving blob, D=downward moving blob. The red circle at 15:24:25 UT shows the collision of a downward moving blob with the top of the flare arcade. An animation of the top and middle panels (AIA 131 \AA) is available. The animation runs from 13:01:37-15:57:25 UT. (An animation of this Figure is available online.) } 
\label{fig8}
\end{figure*}

\section{OBSERVATIONS}\label{obs}
We analyzed SDO/AIA full-disk images of the Sun (field-of-view $\approx$ 1.3~R$_\odot$) with a spatial resolution of 1$\arcsec$.5 (0$\arcsec$.6~pixel$^{-1}$) and a cadence of 12~s, in the following channels: 304~\AA\ (\ion{He}{2}, at temperature $T\approx 0.05$~MK), 171~\AA\ (\ion{Fe}{9}, $T\approx 0.7$~MK), 193~\AA\ (\ion{Fe}{12}, \ion{Fe}{24}, $T\approx  1.2$~MK and $T\approx 20$~MK), 211~\AA\ (\ion{Fe}{14}, $T\approx 2$~MK), AIA 94~\AA\ (\ion{Fe}{10}, \ion{Fe}{18}, $T\approx$1 MK, $T\approx$ 6.3 MK), and 131~\AA\ (\ion{Fe}{8}, \ion{Fe}{21}, \ion{Fe}{23}, i.e., $T\approx$ 0.4, 10, 16 MK) images.
 A noise-gating technique \citep{deforest2017} was used to reduce noise in the SDO/AIA images. We also examined available IRIS slit-jaw images (18-s cadence, 0.16$\arcsec$~pixel$^{-1}$) of the jet source region in the 1400 \AA\ (\ion{Si}{4}, log T (K) = 3.7-5.2) channel. We also utilized simultaneous Hinode Solar Optical Telescope (SOT: \citealt{tsuneta2008}) Ca II H line observation (0.1 $\arcsec$ pixel$^{-1}$) of the source region. STEREO-A observed the same AR on the disk. We used STEREO-A/Extreme UltraViolet Imager (EUVI-A) 195 {\AA} images (10-min cadence) to view the magnetic topology of the active region from a different angle. The separation angle between SDO and STEREO-A was 155$^{\circ}$ on 2014 April 3. The size of the EUVI-A image is 2048$\times$2048 pixels (1.6$\arcsec$ pixel$^{-1}$), covering a field of view out to 1.7 R$_\sun$. 

\subsection{Magnetic configuration and pre-eruption activity}
 To determine the magnetic topology of the source region, we utilized a potential-field extrapolation code \citep{nakagawa1972} from the GX simulator package of SSWIDL \citep{nita2015}. The code was applied to data obtained by SDO's Helioseismic and Magnetic Imager \citep[HMI;][]{scherrer2012} on 2014 March 30 (Figure \ref{fig1}(a)), four days prior to the failed eruption. The potential-field extrapolation (Fig.\ \ref{fig1}b) of AR 12018 reveals the fan-spine topology (cyan color) at the filament eruption site and significant flux (red) overlying the eruption site in AR 12018 (cyan). The northern footpoints of these overlying arcades were connected to the sunspot ($B_{max} \approx -$1200 G) in AR 12017.
AIA 193 and STEREO EUVI-A 195 {\AA} images (Figure \ref{fig1}c,d) show NOAA 12018 from two different viewing angles. AIA covered the eastern part of the active region, while the western part was behind the limb (Figure \ref{fig1}(c)). The IRIS FOV (white box in Figure \ref{fig1}(c)) covered only the southern loops. The STEREO image shows the complete AR 12018, consisting of two filaments F1 and F2 (Figure \ref{fig1}(c, d)). F1 was located along the eastern polarity inversion line (PIL) and F2 lay along the western PIL (marked by white arrows in (c) and (d), respectively). F2 was not visible in the AIA because it was behind the limb.  F1 was not clearly seen in the AIA images, due to blockage by a foreground filament outside the AR, so only the top of the activated filament was clearly visible to AIA. A portion of F1 erupted but failed to produce a CME, whereas F2 remained stationary. 

To determine the 3D coordinates of the identical structures in the AIA and STEREO fields of view, we used the {\it scc\_measure} routine available in SSWIDL. In Figure \ref{fig1}(c) and (d), the southern loop system is marked by blue plus symbols, the red plus signs indicate a pre-eruption jet, and the yellow arrow marks the cusp. Note that the cusp is near the top of filament F1. The apparent position of the cusp beneath part of F1 is due to projection effects. A system of loops (green plus signs) appears north of the AR. We infer that AR 12018 had a fan-spine magnetic geometry (dome width $\approx$ 50$\arcsec$), with a pre-eruptive null about 50-60$\arcsec$ above the photosphere.

During the two hours prior to the slow rise of F1, we observed repeated pre-eruption jets originating from the cusp region (Figure \ref{fig2}(a)).  AIA 211 {\AA} running-difference images and the accompanying movie reveal first jet onset at  $\approx$ 12:35 UT (speed $\approx$ 130 \kms) and at least three faint jets between 13:30-14:00 (marked by arrows in the TD plot, Figure \ref{fig2}(b)). STEREO EUVI-A images also reveal the same narrow jets emanating from the cusp (Figure \ref{fig2}(c-e)). The faint reverse-drift features in the TD plot are coronal rain plasma draining from the top of the cusp, better seen in the AIA 304 {\AA} channel. 

\subsection{Filament activation and slow rise}
The EUVI-A images captured more activity near the cusp after the pre-eruption jets mentioned above. A bright blob was detected north of the cusp (white arrows in Figure \ref{fig3}(a2,a3)) at 14:15:30 UT. Over an hour later, the southern arcade expanded and a small brightening appeared near the cusp at 14:25:30 UT (Figure \ref{fig3}(a4,b1)). 
In the next image in the sequence, a bright loop appeared above filament F2, on the other side of the PIL (Figure \ref{fig3}(b2), marked by an arrow). 
Some of the southern loops opened during this phase, as shown in the running-difference plots (Figure \ref{fig3}(b2) and accompanying animation). A structure above the cusp erupted at 14:35:30 UT (Figure \ref{fig3}b2,b3). 

The Hinode/SOT images and accompanying movie show initial activation of filament F1 at $\approx$ 14:15 UT (marked by arrows in Figure \ref{fig4}(a2,a3)).  The filament disappeared from the SOT field of view at $\approx$ 14:28 UT, before SOT observations ended at 14:30 UT. Coronal rain was also observed along the southern loop system. These dynamic features are visible most clearly in the animation accompanying Figure \ref{fig4}. 

Interestingly, the AIA 193 {\AA} image at 13:57:37 UT (Figure \ref{fig4}(b1)) reveals a likely null point (N) and part of the quasi-circular arcade above F2, located behind the limb. AIA 193 {\AA} running-difference images at around 14:10 UT and the accompanying movie reveal the slow rise of F1. The activated filament, seen in the Hinode/SOT field of view at 14:15 and 14:19 UT, looks like it is a part of F1 (Figure \ref{fig4}(a1,a2)). The accompanying movie shows the rotation of the southern threads (marked by arrows, Figure \ref{fig4}(b2,b3,b4)) of the rising filament. Tracking of the black and white features in the running-difference movie accompanying Figure \ref{fig4} suggests counterclockwise rotation of the rising filament segment. The presence of multiple threads (Figure \ref{fig4}(c3)) indicates a clear connection between the rising F1 and the southern segment of the filament channel. Along with rain and multiple plasma blobs (Figure \ref{fig5}), we see downflows along the southern loops (Figure \ref{fig4}(c2,c3)).

The 193 {\AA} TD intensity plot along slice S1 shows the slow rise ($\approx$ 10 \kms) of filament F1 in absorption during 14:10-14:35 UT. As the filament accelerated from 12 \kms to 62 \kms (14:40-15:00 UT), blobs and downflows appeared (Figure \ref{fig4}(d1)). The filament segment near the cusp (slice S2) rotated (Figure \ref{fig4}(b3)) and some of the cool plasma drained along one leg with $\approx$ 13 \kms (Figure \ref{fig4}(d2)). The southern arcade rose slowly in the interval 14:15-14:42 UT, as revealed by the TD plot of slice S3 (Figure \ref{fig4}(d3)), followed by a sudden rise ($\approx$ 22 \kms) and disappearance of loops during rotation/contraction of the prominence segment. 

\subsection{Blobs and flows during the filament eruption}
Images in AIA 304 {\AA} and hotter channels (131, 171, 193 {\AA}) reveal multiple plasma blobs (marked by arrows in Figure \ref{fig5}) originating near the cusp and travelling along the loops during 14:41-14:53 UT.  IRIS 1400 {\AA} high-resolution images show some of the selected blobs (Figure \ref{fig6}). The nanojets (transient unidirectional jet-like features perpendicular to the loop) reported by \citet{antolin2021} were observed in the same time interval. Most blobs have widths $\approx$ 2-3$\arcsec$ and are accompanied by hot plasma flows along the loops.

AIA 94 {\AA} images exhibit flows and transient heating near the cusp (marked by arrows in Figure \ref{fig5} bottom panels) during 14:25-14:50 UT. The flows from the vicinity of the cusp are consistent with the flows and blob seen in the STEREO EUVI images (Figure \ref{fig3}(a4, b1)), and are more easily seen in the accompanying animations. Heating of the loops below the cusp and downflows along the dome were observed during 14:25-14:53 UT. 

Flaring begins at $\approx$ 14:55 UT with the appearance of multiple blobs near the cusp (see animation accompanying Figure \ref{fig7}), seen in several AIA channels (304, 171, 193, 131 {\AA}). A non-radial plasma sheet structure was observed inside the dome at $\approx$ 15:05 UT onward (Figure \ref{fig8} top panels). Multiple plasma blobs (width $\approx$ 2-3$\arcsec$) were detected along the plasma sheet (marked by arrows in Figure \ref{fig8}) until $\approx$ 15:38 UT. These blobs are similar in size to the blobs seen during the earlier episode of activity (Figure \ref{fig5}). Some of the blobs merge near the cusp and show evidence of heating and/or increased density. Hot flare arcades (FA) appear $\approx$ 15:08 UT onward. A few of the blobs move down (marked by D, Figure \ref{fig8}) along the plasma sheet and collide with the top of the flare arcade (red circle, Figure \ref{fig8}). The plasma sheet disappeared around 15:28 UT. The outflows and some of the blobs near the cusp were also detected in the 94 {\AA} channel, where the cusp appears most clearly (Figure \ref{fig8}).  On the other hand, the multiple bidirectional blobs in the plasma sheet appear most clearly in the 193 {\AA} images (Figure \ref{fig8} bottom panels). 

\begin{figure*}
\centering{
\includegraphics[width=18cm]{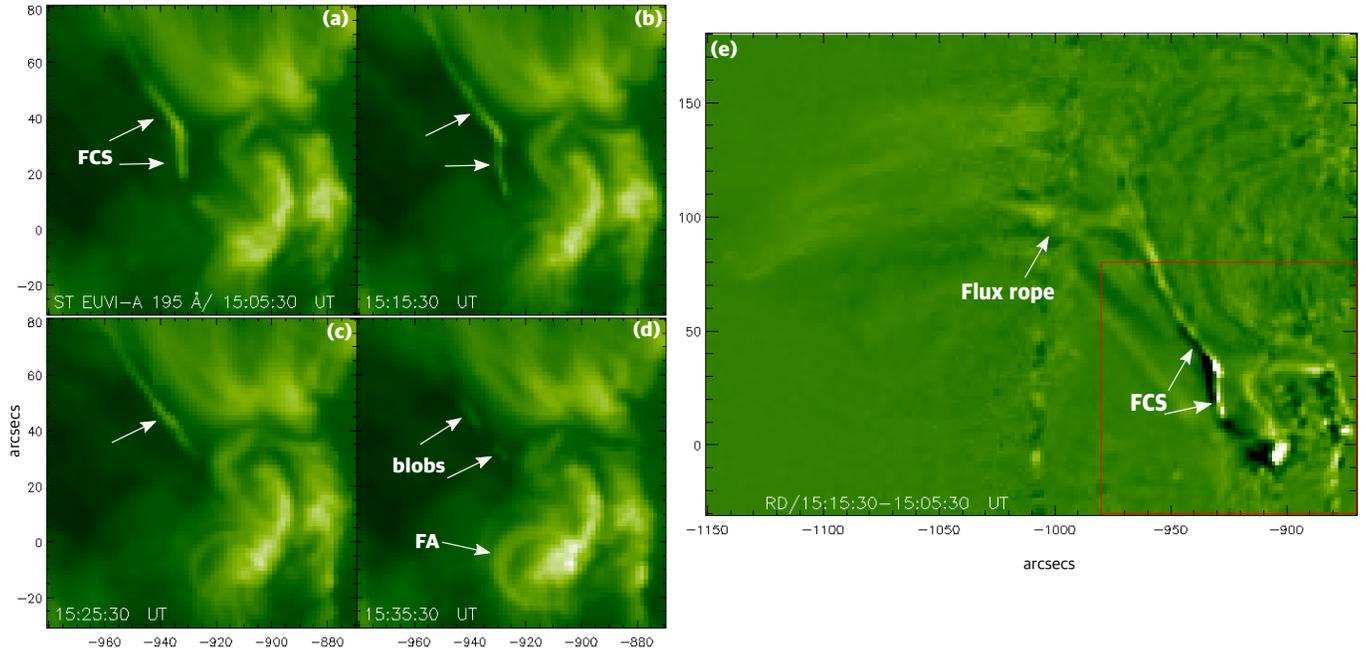}
}
\caption{ (a-d) STEREO/EUVI-A 195 {\AA} images during the second episode of energy release (15:05-15:35 UT). FCS=flare current sheet. FA=flare arcade. (e) EUVI-A 195 {\AA} running-difference ($\Delta t$ = 10 min) image showing the FCS below the erupting plasma. The red box outlines the region shown in panels (a)-(d). An animation of this Figure is available. The animation runs from 12:15:30-16:25:30 UT.  (An animation of this Figure is available online.)} 
\label{fig9}
\end{figure*}

The different point of view of the EUVI-A images, in which the AR is on the disk near the east limb, enables clearer delineation of key features in the region of interest. In particular, EUVI-A 195 {\AA} images reveal a bright plasma sheet (denoted FCS, marked by arrows in Figure \ref{fig9}) that extends $\approx$ 20-25$\arcsec$ below the cusp and up to $\approx$ 40-50$\arcsec$ above the cusp. We also observed some blobs in the vertical FCS (Figure \ref{fig9}(d)). The non-radial plasma sheet below the cusp is the same plasma sheet seen in the AIA 131 {\AA} channel (Figure \ref{fig8}). The extended part of the FCS above the cusp is less clear in the AIA data (due to projection), but the STEREO images confirm the plasma sheet below the rising ejection (Figure \ref{fig9}(e)). 

\begin{figure*}
\centering{
\includegraphics[width=18cm]{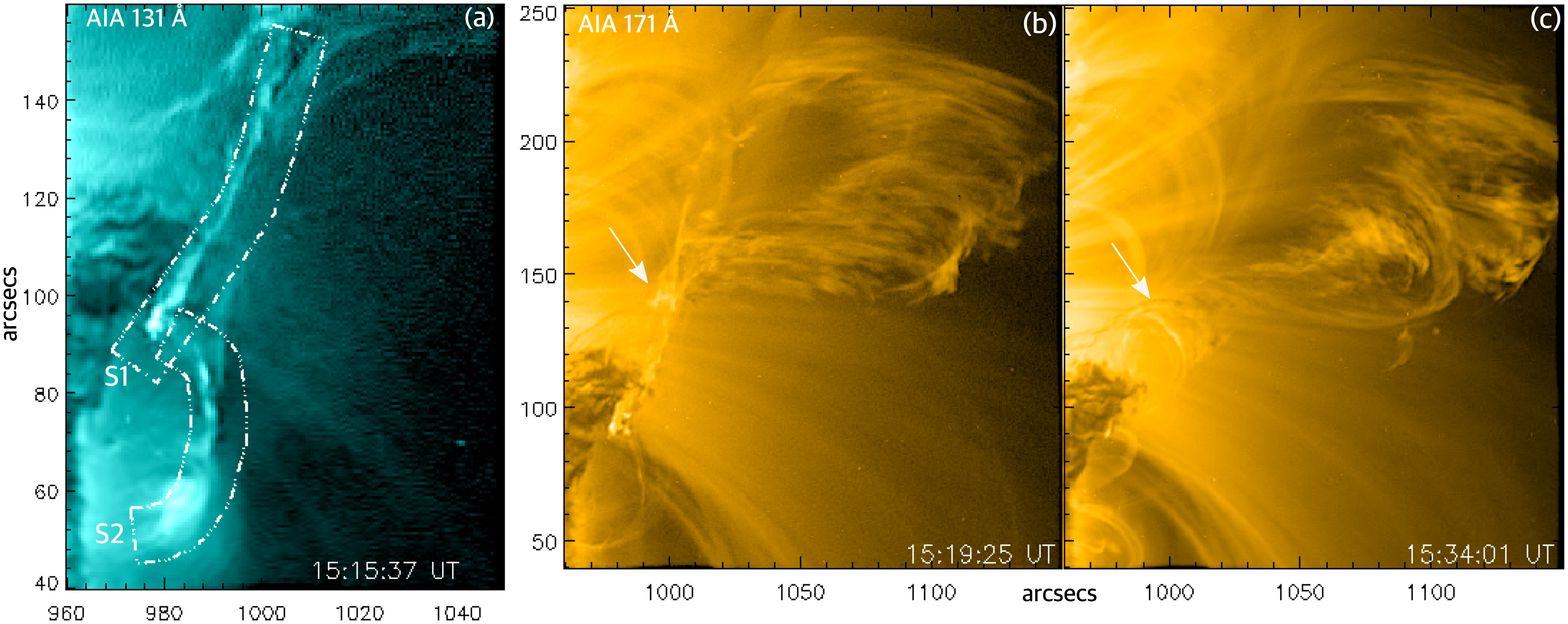}
\includegraphics[width=18cm]{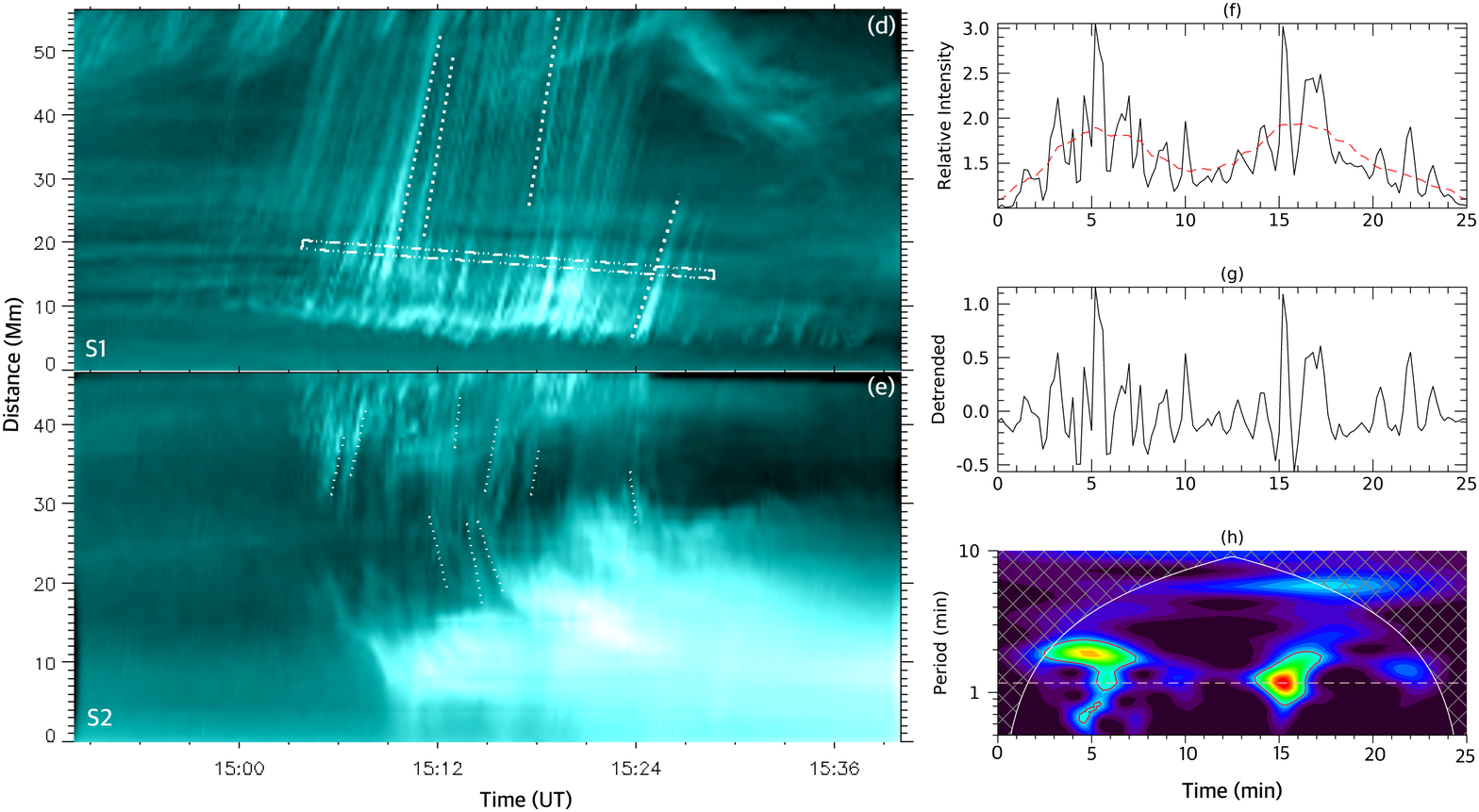}
}
\caption{ (a,b,c) AIA 131 {\AA} and 171 {\AA} images showing the plasma sheet (FCS) below the ejection at three times. S1 and S2 are the slices used to create the time-distance intensity plots. Arrows in panels (b) and (c) indicate the leg-leg interaction and closing afterward forming an arcade.  (d,e) AIA 131 {\AA} time-distance intensity plots along slices S1 and S2. (f,g) Relative intensity (arbitrary unit) profile along the horizontal cut marked in the slice S1 TD map. The red dashed curve is the detrended pattern, which is subtracted from the original signal to obtain a detrended intensity for the wavelet analysis. (h) The wavelet power spectrum of the detrended intensity. The pink horizontal dashed line indicates a period of 70 s. An animation of the panels (a), (d), and (e) is available. The animations run from 14:50:01-15:40:01 UT. (An animation of this Figure is available online.)} 
\label{fig10}
\end{figure*}

To determine the speed of the blobs moving along the plasma sheet, we tracked selected blob paths in the TD intensity maps (Figure \ref{fig10}(d,e)) along two slits outlined in Figure \ref{fig10}(a). Some of the blobs moving upward from the cusp along slit S1 (marked by dotted lines, Figure \ref{fig10}(d)) have speeds around 150-270 \kms. Upward-moving blobs along S2 (marked by dotted lines, Figure \ref{fig10}(e)) in the curved plasma sheet have speeds around 148-285 \kms, whereas the measured downward-moving blobs travel at $\approx$ 98-270 \kms.

The relative intensity of blobs and outflows along the slit shows a quasiperiodic pattern (Figure \ref{fig10}(f-h)). To obtain a detrended light curve for the wavelet analysis, we subtracted the red dashed pattern from the original light curve. A wavelet analysis of the detrended light curve reveals a period $\approx$ 70 s (pink dashed line) with 99\% significance. Similar periods were detected for the upward- and downward-moving blobs in the plasma sheet. 

\subsection{Flare and associated activity}
The GOES soft X-ray (1-8 {\AA}) flux profile revealed a C-class flare during the analyzed time interval, but there was another flare from a different active region nearly at the same time. Therefore, we used the AIA 94 {\AA} channel mean counts of the flare region (shown in Figure \ref{fig5}) as a flare proxy (Figure \ref{fig7}(b)), and avoided using the GOES soft X-ray flux measurements that combined two flares from different ARs. The preflare onset was around 14:42 UT, the impulsive phase began at $\approx$ 15:06 UT, and the maximum occurred $\approx$ 15:30 UT. Therefore the blobs, jets, and downflows along the dome at $t \approx$ 14:40-14:53 UT occurred during the preflare phase, while the bright plasma sheet beneath the flux rope was associated with the flare impulsive phase (15:06-15:28 UT). 

The AIA 304 {\AA} TD intensity plot shows filament activation at $\approx$ 14:15 UT (first vertical dashed line, Figure \ref{fig7}(c)). Jet-like transverse features (marked by J within the slit) were interpreted as nanojets by \citet{antolin2021}. Most of these jet-like features were associated with bright blobs (TD plot during 14:42-14:54 UT) moving through the cool coronal-rain plasma. The erupting prominence segment rotated at the same time (Figure \ref{fig4}(d2)). As seen in several AIA channels (304, 171, 193 {\AA}), the apparent contraction of the dome began at $\approx$ 14:48 UT and became faster ($\approx$ 13 \kms) during 14:53-15:27 UT (Figure \ref{fig7}(c,d)). Brightenings around 20 Mm along the slit from 15:05-15:25 UT in the TD plots are from multiple blobs in the plasma sheet crossing the slit.  

The top of the filament and surrounding hot plasma reached a maximum height of 145 Mm at 15:28 UT, then oscillated in place until 16:12 UT. A separate analysis of these kink oscillations is discussed by \citet{kumar2022}. Thereafter cool plasma drained along the field. No CME was observed in association with this event. \\

\section{INTERPRETATION}
\label{discussion}
 As described above, we analyzed multiwavelength observations from SDO/AIA, IRIS, and STEREO and identified a series of activities associated with a failed filament eruption and a flare. The source was a small active region with a fan-spine topology surmounted by a null point within the observed cusp. When stressed, this type of configuration is well known to produce a wide range of solar eruptions through the breakout mechanism, from tiny jets to eruptive flares \citep{antiochos1999, karpen2012, karpen2017, wyper2017, wyper2018, wyper2021, dahlin2019}.  Our interpretation of the observed features is based on that model. 

According to the breakout scenario, the plasma sheet observed near the cusp encompassed a breakout current sheet (BCS) lying along the fan separatrix, while the plasma sheet observed beneath the rising filament encompassed the flare current sheet (FCS) formed in the wake of the rising filament-channel flux. Pre-eruption jets originating from the null during the two hours prior to the filament slow rise were caused by slow reconnection at the BCS \citep[\eg][]{lynch2013,kumar2021,wyper2021}. Additional evidence for early breakout reconnection was provided by the persistent, strong coronal rain observed well before the eruption. The rain emanated from the vicinity of the null and drained along the envelope of a portion of the fan loops, as reported previously in breakout jet observations \citep{kumar2019b, kumar2021}.  Interchange reconnection has been proposed as a source of coronal rain in pseudostreamers, which also manifest a fan-spine topology \citep{mason2019}.

\begin{figure*}
\centering{
\includegraphics[width=18cm]{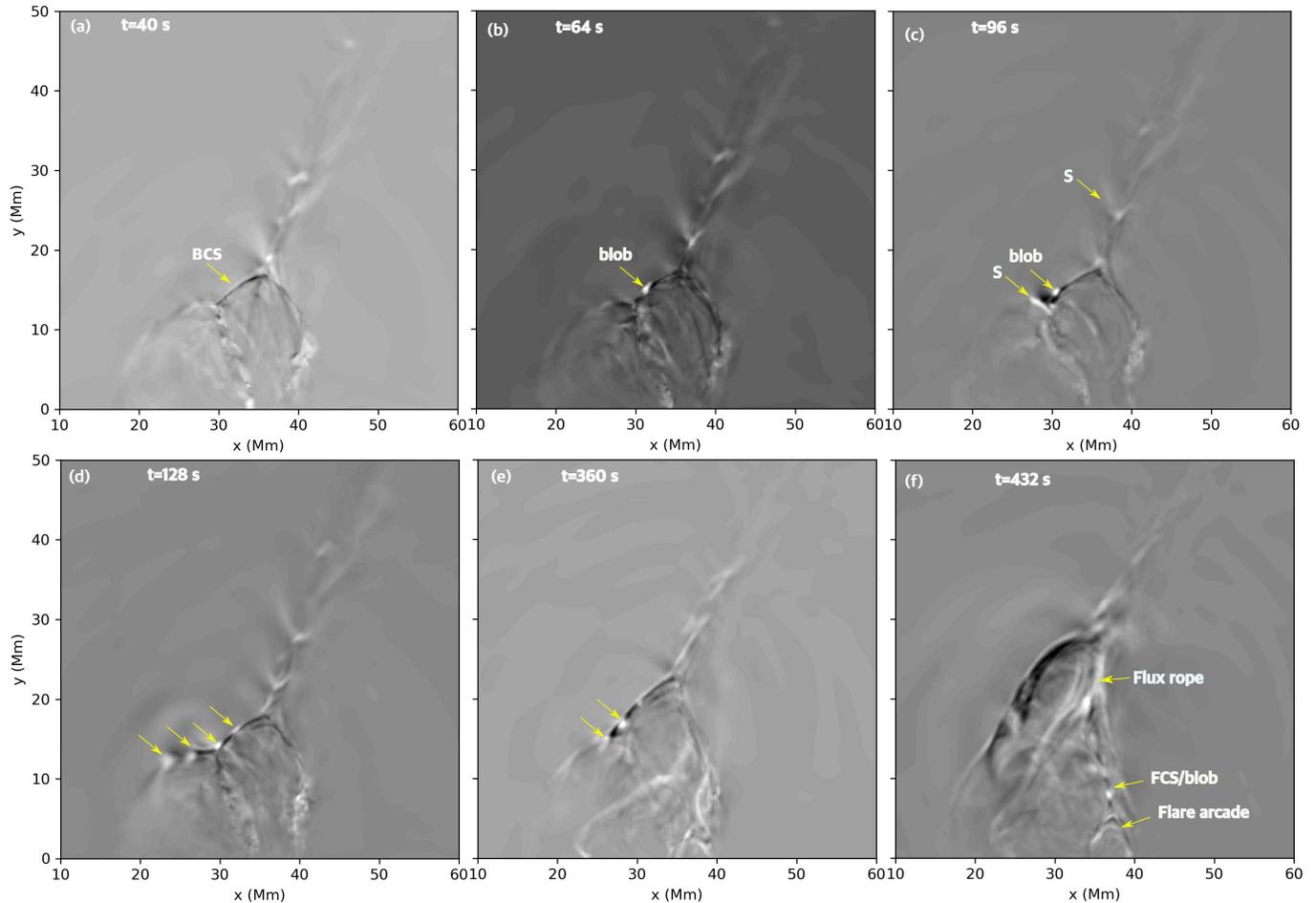}
}
\caption{Line of sight integrated running-difference density images showing formation and propagation of multiple plasmoids (marked by arrows) in the breakout current sheet (BCS) associated with slow rise of filament channel (From \citealt{wyper2018}). FCS=flare current sheet, S=shock front associated with plasmoid. The temporal resolution (cadence) of the simulation is 8 s. (An animation of this figure is available online.)} 
\label{fig11}
\end{figure*}

As breakout reconnection proceeded, the filament began to rise slowly and continued at this pace for approximately 35 min.  The filament and surrounding flux rope remained relatively small.  
We interpret the plasma blobs appearing in the AIA and EUVI images of the cusp region during the preflare phase as plasmoids (magnetic islands) formed by repeated episodes of reconnection in the BCS and subsequently released onto the fan loops.  In 3D, plasmoids are tiny flux ropes formed by multiple reconnections in current sheets \citep{guidoni2016,dahlin2019,lynch2020}. For comparison, Figure \ref{fig11} shows results of a high-resolution MHD simulation of a coronal jet by \citet{wyper2018}, which demonstrates formation and propagation of multiple plasmoids (marked by arrows) in the BCS during the slow rise of the filament channel flux (density images in Figure \ref{fig11}). When these plasmoids exit the BCS, their magnetic structures lose their integrity through reconnection with the unidirectional fan field. The density enhancements remain, however, and propagate toward the footpoints (Fig. \ref{fig11}(d)-(f)). 

The early flare phase was characterized by jets and downflows along the separatrix, heating of the southern fan loops, and the fast rise of the filament and surrounding flux rope. We interpret these features as key signatures of the onset of flare reconnection, which also increases the reconnection rate at the BCS.  Note that the blobs/jets reported by \citet{antolin2021} were observed during this period, but their analysis was based primarily on IRIS observations of the southern loops and did not include the filament eruption and flare.  Based on the simulation results outlined above, we conclude that the transverse jet-like features associated with blobs near the cusp in our event may arise due to component reconnection \citep{swisdak2005, fuselier2011} between BCS plasmoids and the ambient fan field. Further study of the interactions between these propagating plasmoids and the ambient field is needed to verify this scenario.

An alternative explanation of the transverse jet-like features comes from our high-resolution MHD simulations of CME onset via breakout reconnection  \citep[\eg][]{karpen2012,guidoni2016}. In these simulations, transverse current and density enhancements associated with bright plasmoids moving along the breakout and flare current sheets are simply signatures of compressive bow shocks driven by the plasmoids. Such shocks are likely to appear bright in EUV images, but it is unclear whether they could persist beyond the current sheets of origin.  

The flare main phase began with the appearance of multiple plasmoids and jets near the cusp. As the filament accelerated, two new features appeared: a bright, linear plasma sheet beneath it, attributed to the lengthening FCS, and hot plasma entwined with the rising and rotating filament that we interpret as mass residing in the flux rope created by reconnection in this FCS. Plasmoids moved upward and downward in the FCS throughout the eruptive phase, and a bright, hot flare arcade appeared within the dome in the southern portion of the AR. The sunward plasmoids collide and merge with the flare arcade, while the upward plasmoids augment the rising flux rope. 

A curved plasma sheet was observed inside the dome during this phase, with the northern end apparently intersecting with the lower end of the linear FCS. We interpret this feature as an extension of the FCS. Multiple plasmoids were detected along the curved plasma sheet, some of which merge near the cusp and show evidence of heating and/or increased density.  A few of the plasmoids move down along the plasma sheet and collide with the top of the flare arcade. Numerical simulations of breakout CMEs/eruptive flares and jets confirm the above sequence of activities during the main flare phase, including the generation and evolution of bidirectional plasmoids in the FCS \citep[\eg ][]{karpen2012,wyper2018}. The blob properties are consistent with those of plasmoids observed in other eruptive events \citep{kumar2013, kumar2019b}. Similar flare current sheets with multiple plasmoids have also been reported in previous observations of flare reconnection below a flux rope \citep{takasao2012,kumar2018,kumar2019b,kumar2019a,kumar2021}. Therefore, we conclude that these blobs are associated with reconnection in the breakout and flare current sheets rather than nanoflares \citep{antolin2021}.

The 193 {\AA} animation accompanying Figure \ref{fig4}(b1-c3) reveals the kinking of the filament and its supporting flux rope, as well as possible leg-leg reconnection (Figure \ref{fig10}(b)) that forms an arcade underneath this reconnection site (Figure \ref{fig10}(c)).  STEREO images clearly show a vertical current sheet with multiple plasmoids below the rising flux rope (Figure \ref{fig9}(e)).  Leg-leg reconnection was predicted by an MHD simulation of a kinking flux rope \citep{kliem2010} and previously inferred from observations of other eruptions \citep{alexander2006,cho2009,kumar2014}.

Quasiperiodic pulsations (QPPs) detected during solar flares have been ascribed to two main mechanisms: MHD waves and bursty reconnection with plasmoids \citep{nakariakov2009,Zimovets2021}. 
 Our observations provide strong evidence that bursty reconnection can produce high-frequency (70-s) QPPs in EUV emissions. Previously, we detected 2.5-3.2 min periodicity in a pseudostreamer FCS with multiple plasmoids \citep{kumar2021}, and 70-s and 140-s QPPs in radio, X-ray, and EUV emissions associated with fast-mode waves (without plasmoids) in a flaring active region \citep{kumar2017}. Therefore, both QPP drivers (MHD waves and plasmoids) have been observed in null-point topologies.

We observed progressive removal of the southern loops (inside the dome) starting around 14:53 UT, and found a clear correlation between the apparent contraction and disappearance of fan loops in the AIA 193/171 {\AA} channel. In addition, the cotemporal EUVI-A 195 {\AA} images show an increasing volume of bright loops over F2 (see Figure \ref{fig9} and accompanying animation). The apparent contraction was detected well above the curved FCS where bidirectional plasmoids were formed intermittently. The apparent contraction of the dome during the event is likely 
due to fast breakout reconnection accelerated by explosive flare reconnection at the FCS. To the best of our knowledge, this type of contraction in an erupting null-point topology has not been reported before. 

Part of the contracting structure in the TD plot (Figure \ref{fig7}) shows associated blobs in the FCS and clearly correlates with the flare energy release. If we assume that the observed contraction speed equals the reconnection inflow speed ( $\approx$ 13 \kms), and the plasmoid speed (100-285 \kms) equals the outflow speed, the estimated reconnection rate (inflow/outflow) in the FCS is in the range 0.05 - 0.13. This reconnection rate is consistent with estimates from previous observations of flare reconnection associated with plasmoids  \citep{takasao2012,kumar2013}. We conclude that the removal via breakout reconnection of closed flux below the separatrix surface reduced the volume of this portion of the fan, and transferred that flux to the portion of the fan overlying F2. Further 3D MHD simulations are required to validate this scenario.

\section{SUMMARY AND CONCLUSIONS}
\label{summary}

We have demonstrated that multiple plasma blobs and jets were detected during a failed filament eruption on 2014 April 3 in NOAA 12018, which had a fan-spine magnetic topology. These blobs are interpreted as multiple magnetic islands (plasmoids) formed by magnetic reconnection in breakout and flare current sheets. 
Breakout reconnection at the breakout sheet is the most likely candidate for the origin of the plasmoids and flows guided along the separatrix, as detected in our previous observations of coronal jets \citep{kumar2019b} and numerical MHD simulations \citep{wyper2016}. We also observed repetitive bidirectional plasmoids (width $\approx$ 2-3$\arcsec$) in the flare current sheet, with a period $\approx$ 70 s during a 25-30 minute interval. The speeds of upward and downward moving plasmoids were 150-285 \kms and 100-270 \kms, respectively. 

The Interpretation section is aimed at explaining our observations of this complex event in terms of a single, compelling eruption model. The breakout model provides a comprehensive accounting of the observations described (see Section 3). We are unaware of any alternative model that can explain all of the observed phenomena. The eruption clearly undergoes low coronal breakout reconnection at the stressed null, and drives enough flare reconnection to form a significant flux rope. The erupting flux rope undergoes kinking but does not escape. 

A major mystery remains: why did this filament eruption fail? We speculate that the flux rope failed to overcome the strapping forces of the overlying large-scale arcades for one or both of the following reasons.  First, breakout reconnection between the leading edge of the rising flux rope and overlying arcades would be affected by the rotation of the flux rope during kinking, which can change the direction of the field at the top of the rope. If the magnetic field of the rotated flux rope and that of the overlying arcades became roughly parallel, the breakout reconnection would diminish. Second, the free magnetic energy transferred from the filament channel to the flux rope could be insufficient to overcome the strapping forces provided by the overlying arcade field. Targeted numerical experiments are needed to clarify the circumstances distinguishing successful from unsuccessful eruptions.

We conclude that plasmoids, associated jets, and fan-loop heating (including a flare arcade) in this event are all due to reconnection during the filament eruption rather than nanoflaring. Future high-resolution observations from DKIST and Solar Orbiter are expected to shed more light on dynamic fine-scale features in eruptions, such as reconnection jets and plasmoids, and provide greater insight into the causes of failed eruptions. 

\acknowledgments
We are grateful to the referee for insightful comments that
have improved this paper. SDO is a mission for NASA's Living With a Star (LWS) program. IRIS is a NASA Small Explorer mission developed and operated by LMSAL with mission operations executed at NASA Ames Research Center and major contributions to downlink communications funded by ESA and the Norwegian Space Centre. STEREO is the mission in NASA's Solar Terrestrial Probes program. This research was supported by NASA’s Internal Scientist Funding Model (ISFM) work package, Guest Investigator (\#80NSSC20K0265), and Living with a Star Strategic Capability (\#80NSSC22K0892) programs. Wavelet software was provided by C. Torrence and G. Compo, and is available at http://paos.colorado.edu/research/wavelets/.

\bibliographystyle{aasjournal}
\bibliography{reference.bib}

{\small

\begin{longtable*}{c c c }
\caption{Time history of the event } \\
\hline \\
\label{tab1}
Time (UT)  &Activity                &Observations                       \\

    \hline
 &                                              &                                                                                        \\  
12:30, 13:15       &Pre-eruption jets                        &AIA and STEREO  \\    
  
            &                                  &                                                                                                \\
            
14:05, 14:15        &Filament F1 slow rise              &Observed partially in high-resolution SOT$^\dagger$ and IRIS$^\star$ images,  \\   
                              &                                  &      complete eruption covered in AIA and STEREO                                                                                           \\                                                            
           &                               &                                                                                                   \\

14:41-14:53      &Blobs during the filament eruption,  Preflare phase,   &AIA, IRIS, STEREO \\
                          & loops expansion/disappearance in AIA 171/193 {\AA},                                &                                                                                   \\    
                             &flows near the null in AIA 94 {\AA}, dome/fan contraction.  & \\       & &                                                                                                                                                        \\    
      
        &                                 &                                                                                   \\    
     14:55-15:38         &Multiple blobs and flows in the flare plasma sheet, filament fast rise,   & AIA, STEREO  \\  

  &  (P $\approx$ 70 s) during the filament eruption                               &                                                                                   \\   
    &                                 &                                                                                                \\ 
 \hline

\end{longtable*}
${}^\dagger$ Hinode SOT observation ends at 14:30 UT.\\
${}^\star$ IRIS Observation ends at 14:53 UT. \\
}


\end{document}